\documentclass[epj,english,twocolumn]{webofc}
\usepackage{graphicx}
\usepackage{subcaption}
\usepackage[varg]{txfonts}   
\usepackage{hyperref}        
\woctitle{Heavy Ion Accelerator Symposium 2019}
\begin{document}
\title{SABRE and the Stawell Underground Physics Laboratory}
%
%
\subtitle{Dark Matter Research at the Australian National University}

\author{L.~J.~Bignell\inst{1}\fnsep\thanks{\email{lindsey.bignell@anu.edu.au}} \and
        E.~Barberio \inst{2} \and 
        M.~B.~Froehlich \inst{1} \and
        G.~J.~Lane\inst{1} \and
        O.~Lennon\inst{1} \and
        I.~Mahmood \inst{2} \and
        F.~Nuti \inst{2} \and
        M.~S.~Rahman \inst{3} \and
        C.~Simenel \inst{1} \and
        N.~J.~Spinks \inst{1} \and
        A.~E.~Stuchbery \inst{1} \and
        H.~Timmers \inst{3} \and
        A.~Wallner \inst{1} \and
        L.~Wang \inst{1} \and
        J.~Wu \inst{1} \and
        Y.~Y.~Zhong \inst{1}
}

\institute{Department of Nuclear Physics, The Australian National University, Canberra, ACT 2601, Australia
\and
           School of Physics, The University of Melbourne, Melbourne, VIC 3010, Australia
\and
           School of Science, The University of New South Wales, UNSW Canberra, Canberra, ACT 2612, Australia
          }

\abstract{%
The direct detection of dark matter is a key problem in astroparticle physics that generally requires the use of deep-underground laboratories for a low-background environment where the rare signals from dark matter interactions can be observed. This work reports on the Stawell Underground Physics Laboratory -- currently under construction and the first such laboratory in the Southern Hemisphere -- and the associated research program. A particular focus will be given to ANU's contribution to SABRE, a NaI:Tl dark matter, direct detection experiment that aims to confirm or refute the long-standing DAMA result. Preliminary measurements of the NaI:Tl quenching factor and characterisation of the SABRE liquid scintillator veto are reported.
}
\maketitle
\section{Introduction}
\label{intro}
Understanding the nature of dark matter is a major goal of modern physics. There is a distinct possibility that dark matter is an as-yet undiscovered fundamental particle; there are several avenues being explored that target the direct detection of particle dark matter \cite{Klasen_2015}. Ultra-low background, underground direct detection searches are a prominent example. Of these, most experiments search for an excess of interaction events beyond known backgrounds in an active detector volume, that may be due to galactic dark matter.

To unambiguously attribute any measured excess to dark matter, an astrophysical signature will be required. For non-directional terrestrial experiments, one such signature is the annual modulation of the rate of dark-matter induced events in the detector, owing to the orbit of the earth around the sun \cite{Freese_2013}. However, any unexplained signal excess that modulates annually could also be due to a poorly understood seasonal systematic effect. Indeed, the DAMA experiment has observed a modulation which could be attributed to dark matter \cite{Bernabei2018}, but their result remains contested due to tension with other experiments and suggested seasonal systematics \cite{Davis_2014, mckinsey2018dama}. There are a number of NaI:Tl based detectors currently operating, including COSINE and ANAIS, that have recently reported their first annual modulation searches~\cite{Adhikari_2019_modulation, Amare_2019_modulation}; in both cases they are not yet sensitive enough to confirm or refute the DAMA result. In addition, both of these detectors are located in the Northern Hemisphere, whereas a measurement in both hemispheres can control for seasonal effects to verify the astrophysical origin of any modulation in a DAMA-like experiment. This is a major goal for the present SABRE experiment, as discussed in Section~\ref{sabre}.

For detectors with a directional capability, there is the possibility to measure an additional diurnal modulation due to the rotation of the Earth, which modulates the detector orientation with respect to the average incident dark matter velocity \cite{Ohare2015}. The magnitude of this modulation depends upon the latitude of the detector. A southern-hemisphere experiment with a latitude equal to Stawell, Victoria has a larger diurnal modulation than any existing underground laboratory~\cite{urquijo2016southern}. Since it does not possess a directional capability we do not expect to observe the diurnal modulation with the SABRE detector \cite{BernabeiConf2018}.

\section{The Stawell Underground Physics Laboratory and the Australian Research Council (ARC) Centre of Excellence for Dark Matter Particle Physics}\label{supl}
The Stawell Underground Physics Laboratory (SUPL) is a new, general-purpose, underground laboratory currently under construction, and is the first such laboratory in the Southern Hemisphere. SUPL is situated 1025 m underground in the Stawell Gold Mine in regional Victoria, Australia at a latitude and longitude of (-36.060, 142.801). The laboratory has similar overburden (2900 m water equivalent) and radioactive background to the Gran Sasso National Laboratory (LNGS). The laboratory design includes a radon suppression system and a clean area. Details on the SUPL design are given elsewhere \cite{urquijo2016southern}.

The creation of SUPL adds a major new capability to Australia's research infrastructure. Furthermore, it will be exploited through the ARC Centre of Excellence for Dark Matter Particle Physics that will support a broad experimental program to leverage SUPL and other Australian areas of strength, that are relevant to dark matter particle physics, in a coordinated way. The Centre will be led by the University of Melbourne, who will work with the Australian National University, University of Adelaide, University of Sydney, University of Western Australia and the Swinburne University of Technology, as well as a large number of Australian and international partner organisations. The Centre will prosecute research across four main themes: direct detection searches for dark matter with underground experiments at SUPL and above-ground experiments targeting axion-like particles; precision metrology to facilitate the ultra-pure, ultra-sensitive experimental measurement program; searches for dark matter production at the Large Hadron Collider; and a theory program focused on dark matter phenomenology. Details of the new Centre are given on its website~\cite{CoEDMPP}.

\section{SABRE}\label{sabre}
An important question facing experimental searches for dark matter is the origin of the signal observed by the DAMA experiment and claimed by them to be direct detection of dark matter. DAMA has measured an annual modulation in the rate of 2-6 keV events in their NaI:Tl detector for many years \cite{Bernabei2018}, which may be interpreted as being due to WIMP dark matter. Given the null results from other experiments (see summaries, for example, in Refs.~\cite{Klasen_2015,PDG2019}), DAMA's measured modulation is strongly inconsistent with most WIMP interaction models. However, the DAMA result has yet to be verified or refuted in a model-independent way with an identical target material.

Given the potential significance of the DAMA measurement, it is essential to verify the observed signal in a separate experiment. 
SABRE is a NaI:Tl-based experiment that aims to conclusively determine whether the DAMA observation originates from dark matter or an experimental systematic effect.

SABRE will consist of two detectors, one located in Gran Sasso National Laboratory (LNGS), Italy and the other in the Stawell Underground Physics Laboratory (SUPL), Australia. Having detectors in the Northern and Southern Hemispheres permits SABRE to control for seasonal systematic effects and provides a powerful means to verify whether the observed modulation has a galactic origin.

\begin{figure}[!tb]
\centering
\includegraphics[width=5cm]{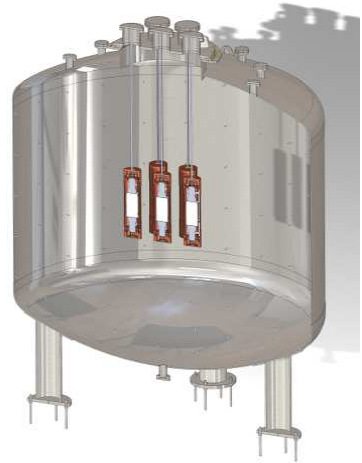}
\caption{A rendering of the SABRE South detector. The central NaI:Tl detectors are surrounded by a vessel filled with liquid scintillator, while the vessel itself will be surrounded by low-radioactivity passive shielding (not shown).}
\label{fig:SSDrawing}       
\end{figure}

Each SABRE detector will consist of 50 kg of ultrapure NaI:Tl contained within a liquid scintillator veto and low-radioactivity passive shielding (Fig.~\ref{fig:SSDrawing}). The full SABRE design and expected sensitivity is outlined in Refs.~\cite{Antonello_2019_MC, Antonello_2019_PoP}, and is expected to confirm or refute the DAMA result at 5$\sigma$ significance within 3 years of operation.

\subsection{Quenching Factor Measurements}\label{QF}
Many WIMP models predict that the most prevalent interaction with galactic dark matter is via coherent elastic nuclear scattering. Thus, it is essential to understand a dark matter detector's nuclear recoil energy scale to interpret a possible signal \cite{Collar2010}. For scintillation detectors, the nuclear recoil light yield is reduced with respect to an electron recoil of the same energy, due to losses to lattice excitations and differences in the electronic stopping power \cite{BIRKS1964}. This can be quantified via the quenching factor; the ratio of the light yields from a nuclear recoil compared to an electron recoil with an identical energy.

While the NaI:Tl quenching factor has been measured previously \cite{Collar2013, Xu2015, stiegler2017QF, joo2018quenching}, there is tension between these measurements that exceeds the experimental uncertainties. The origin of the difference in the reported values is unclear, as different methods have been used and possible causes of systematic effects 
were not always reported. The DAMA experiment has assumed an energy-independent Na quenching factor of 0.3 for their analysis, which is also at odds with the most recent quenching factor measurements.

In preparation for investigating the quenching factor of the SABRE crystals, we have performed measurements using an existing NaI:Tl crystal originally produced in the USSR circa 1988. Its dimensions were 40 mm high by 40 mm diameter and the Tl doping concentration was not specified on the datasheet. Two high quantum efficiency 1"x1" photomultipliers (Hamamatsu H11934-200, peak Q.E. of 43\%) were coupled to the single quartz window for readout. The low-energy, electron-recoil calibration used $^{241}$Am, $^{137}$Cs, and $^{133}$Ba sources. The calibration was corrected for the known nonlinear NaI:Tl electron recoil response~\cite{ValentineAndRooney1998, Khodyuk_2010}.

Nuclear recoils were generated by elastic neutron scattering at the ANU 14UD Heavy Ion Accelerator. A pulsed proton beam (<2~ns on, 749~ns off) with energies of 3 MeV, 5.2 MeV, and 6 MeV was used to irradiate a 500~$\mu$g/cm$^{2}$ LiF target backed with sufficient Ta to stop the beam. Quasi-monoenergetic neutrons produced by the $^{7}$Li(p,n) reaction were collimated onto the NaI:Tl crystal using a polyethylene collimator. Six EJ-309 liquid scintillator detectors were located at 12$^{\circ}$, 22.5$^{\circ}$, 40$^{\circ}$, 67.5$^{\circ}$, 112.5$^{\circ}$, and 135$^{\circ}$ with respect to the beam axis. These were placed at various distances ranging from 15 to 150~cm away from the NaI:Tl detector, with the closest having an angular acceptance of 0.2 Sr. Waveforms in the NaI:Tl and liquid scintillator detectors were acquired using a PIXIE16 digitizer from XIA~\cite{XIA}, sampling at 500 MSPS with 12 bit resolution. The waveform capture was gated on coincidences between the NaI:Tl and any of the liquid scintillator detectors. The beam timing signal from the accelerator was also recorded to facilitate time-of-flight calculations.

The digitized waveforms were processed to estimate the pulse arrival time, charge, and (for the liquid scintillation detectors) a particle identification metric. Baseline noise was removed using a Wiener filter \cite{smith2003DSP} built using the averaged single photoelectron waveform as the signal template. The filtering greatly improved the pulse arrival time estimate, which was determined using a threshold. The pulse charge was calculated by integrating the unfiltered waveforms. However, to avoid integrating noise on the NaI:Tl waveforms in particular, the charge was evaluated for regions that exceeded a threshold, including 4~ns on either side of these regions. For the liquid scintillation detectors, the charge was evaluated in two fixed integration windows relative to the inferred pulse arrival time. The ratio of the charges in these windows was adopted as the particle identification metric. The window lengths were chosen to maximise a figure of merit given by $\frac{\mu_{n} - \mu_{\gamma}}{\sigma_{n} + \sigma_{\gamma}}$, where $\mu$ and $\sigma$ are the mean and standard deviation of peaks in the particle identification metric corresponding to the neutron- and gamma-induced events.

\begin{figure}
    \centering
    \includegraphics[width=8.1cm]{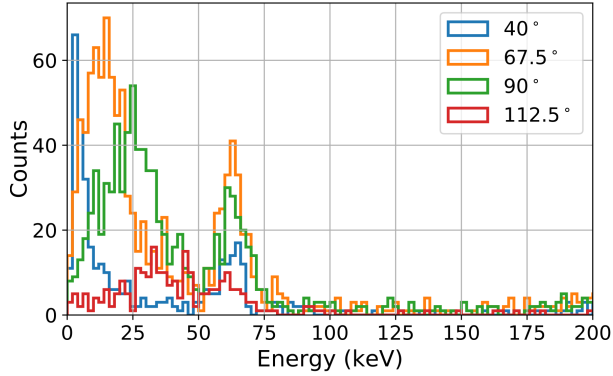}
    \caption{Measured nuclear recoil spectra, after applying the analysis cuts, for detectors at various angles and with a 3 MeV proton beam.}
    \label{fig:RecoilSpec}
\end{figure}

Analysis cuts were applied so as to select nuclear recoil events in the NaI:Tl detector that are a result of elastic neutron scattering.  The times-of-flight between the pulsed beam arrival at the target and signals occurring in both the NaI:Tl and the liquid scintillator, were required to be consistent with a neutron, while the particle-identification metric measured by the liquid scintillator detector was gated to select neutron-like events. 

Example recoil distributions after applying cuts are shown in Fig.~\ref{fig:RecoilSpec}. There are peaks in the spectra that can be attributed to nuclear recoils from Na and I and have the expected dependence upon recoil angle. The peak at $\sim$60~keV that is present in all spectra is from the inelastic excitation of $^{127}$I. The recoil peaks were fit by a Gaussian with a linear plus constant background. The incident neutron energy was calculated using PINO \cite{Reifarth_2009}. The recoil-energy calculation used the mean angle of each liquid scintillation detector with respect to the beam. A full Monte Carlo treatment of the expected distribution of nuclear recoil events in the NaI:Tl is underway, and will be reported in a future publication.

\begin{figure}
    \centering
    \includegraphics[width=8.1cm]{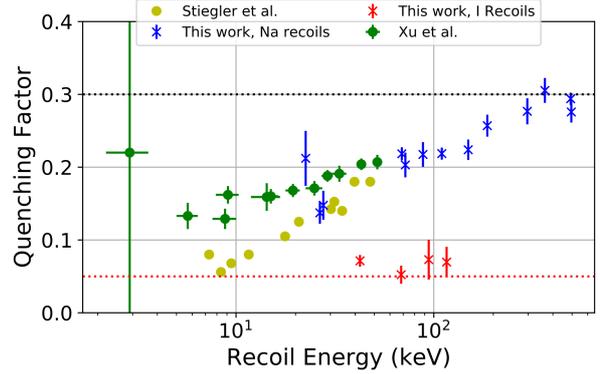}
    \caption{The quenching factors for Na and I recoils in NaI:Tl measured both in this work and in  previous measurements \cite{Xu2015, stiegler2017QF} taken with a similar methodology to that used here. The black and red dashed lines show the constant quenching factor values adopted by DAMA for Na and I, respectively.}
    \label{fig:QFs}
\end{figure}

The quenching factors determined from these measurements are given in Fig.~\ref{fig:QFs}, along with values from two other publications~\cite{Xu2015, stiegler2017QF}. The present results are in agreement with the high-energy extrapolation of both of these prior datasets, but do not yet distiniguish between them at the lower energies where the DAMA signal has been observed. The present 1980s detector has provided an effective proof-of-principle and it is anticipated that future measurements with a SABRE crystal (cut to a 1-inch cube to specifically match the high QE PMTs) will result in a low-energy threshold that probes quenching factors at the low energies required.

\subsection{Veto Scintillator Characterisation}
\label{sec:LS}
The SABRE veto functions both as high-purity active shielding and as a means of eliminating decay events from intrinsic radioimpurities in the inner detector volume that have correlated gamma-ray emissions. SABRE North will use a pseudocumene-based liquid scintillator from the Borexino experiment \cite{Benziger_2008}, which is situated adjacent to SABRE at LNGS, and has well-understood properties. Due to operational constraints associated with handling flammable liquids in a mine environment, SABRE South will not use pseudocumene, and will instead use a linear alkylbenzene (LAB) based scintillator. While the final fluorophore concentration is being optimised for the experiment, 2,6-diphenyl oxazole (PPO) will be used as the primary fluorophore, and 1,4-Bis~(2-methylstyryl)~benzene (bis-MSB) being evaluated as a secondary fluorophore. We do not expect that the use of different veto materials across SABRE North and South will significantly affect our sensitivity to annual modulations in two hemispheres.

To facilitate characterisation studies of LAB scintillator, we have purchased a bulk quantity (1 m$^{3}$) of industrial grade LAB (Jintung Petrochemical) and developed a simple purification process using vacuum distillation. Acceptable distillation results were achieved with a temperature of 120 $^\circ$C and a pressure of 0.05 mBar. The important material parameters from a particle detection point of view are the scintillation light yield and the optical attenuation length; measurements of these are presented below. 

\begin{figure}
    \centering
    \includegraphics[width=8cm]{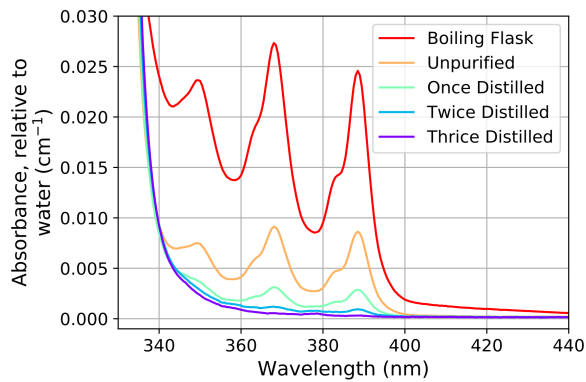}
    \caption{The optical absorbance of linear alkylbenzene at various levels of purification via vacuum distillation. ``Boiling Flask'' refers to the remnant material that remains after distillation.}
    \label{fig:UVVISimpurities}
\end{figure}

\subsubsection{Optical Absorbance}
Measurements of the optical absorbance are achieved using a dual-beam UV-VIS spectrophotometer (Shimadzu UV-2600), and are performed relative to deionised water. Several unknown optical impurities are visible in the absorbance spectrum of unpurified LAB, and measurements of samples taken from the distillate and boiling flask confirm that these are less volatile than the LAB (see Fig.~\ref{fig:UVVISimpurities}).

\subsubsection{Light Yield}
\begin{figure}[h]
 
\begin{subfigure}[b]{2cm}
\includegraphics[width=2cm]{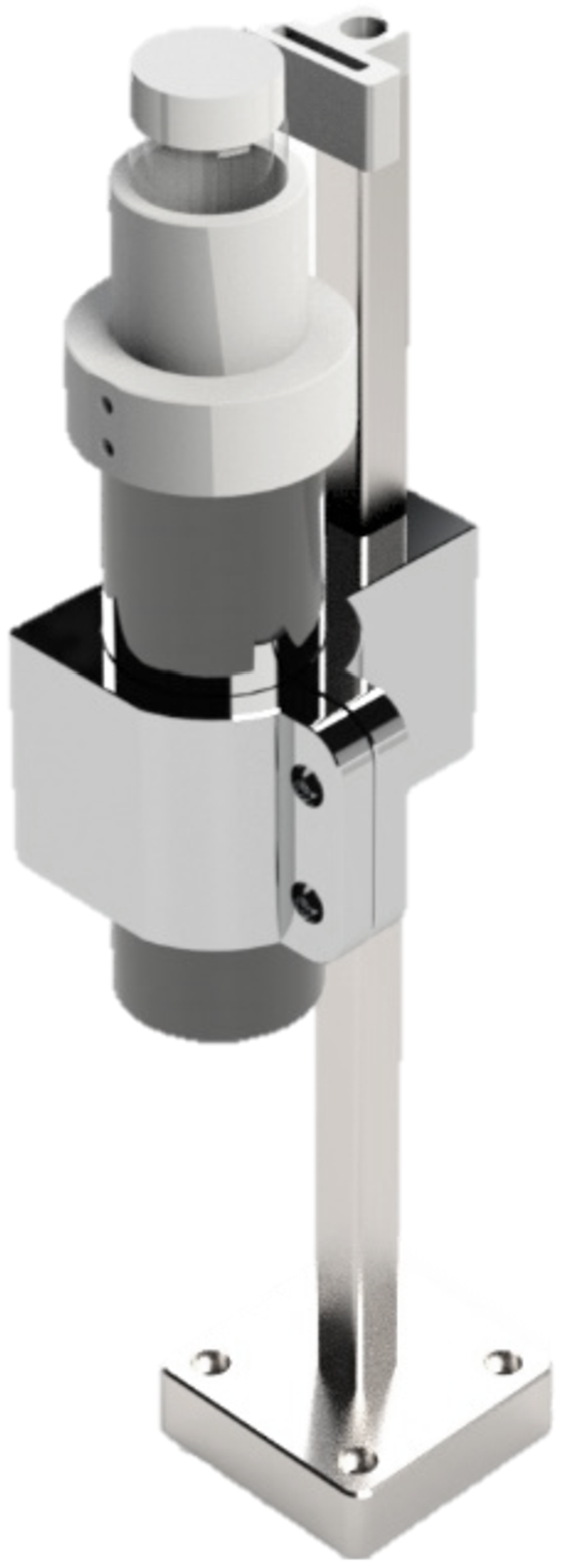} 
\caption{}
\label{fig:LYsetup}
\end{subfigure}
\begin{subfigure}[b]{6.4cm}
\includegraphics[width=6.4cm]{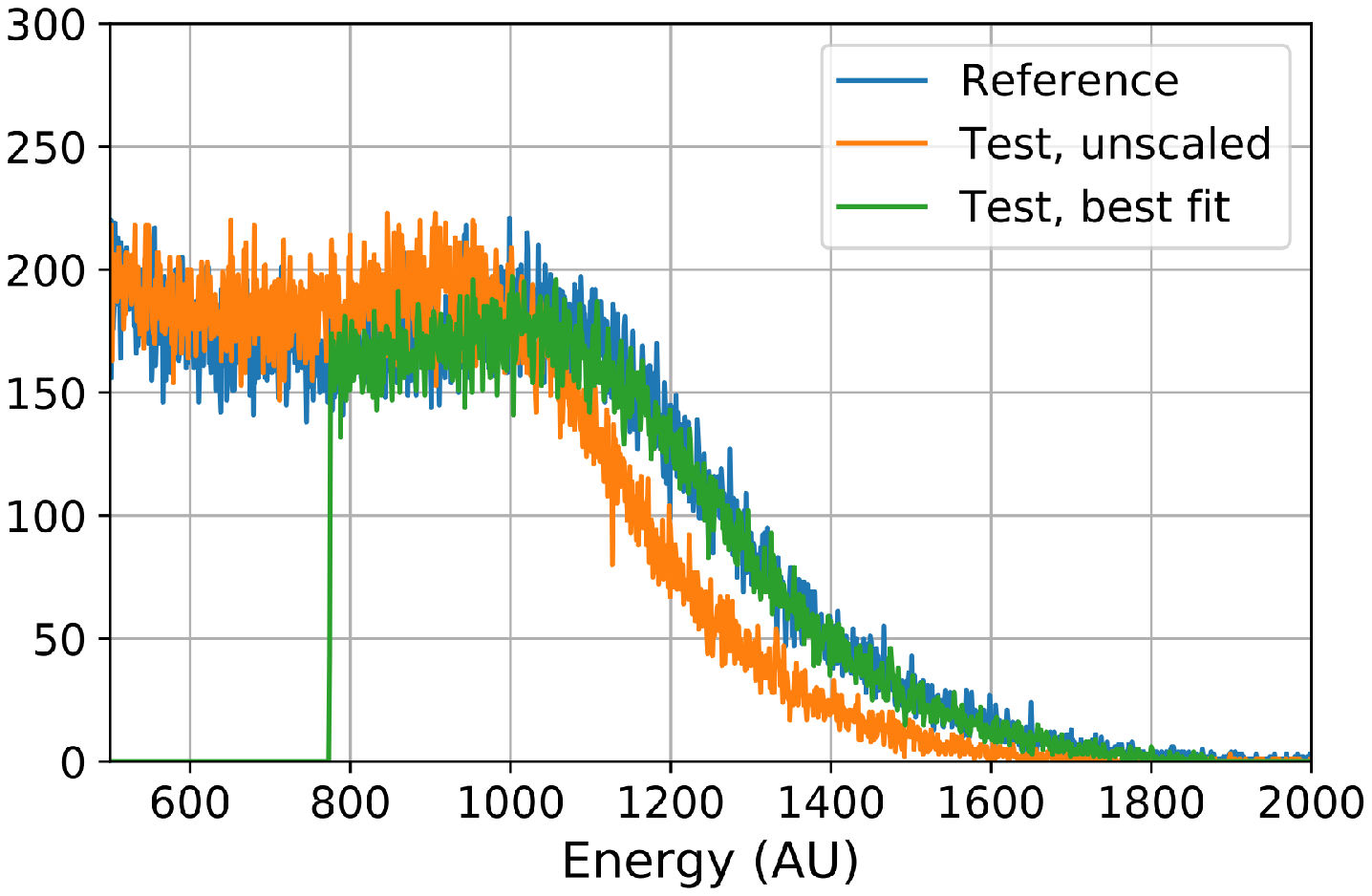}
\caption{}
\label{fig:LYdemo}
\end{subfigure}
\caption{(a) Rendering of the light yield test setup that is contained in a dark box. The lid of the vial is shown, while the $^{137}$Cs source is placed in the holder (small slit) atop the mounting post. (b) A reference spectrum (blue) is compared with a test spectrum (orange) over a limited range with various scaling factors. The green trace represents the rescaled test spectrum with minimum $\chi^2$ that best reproduces the blue spectrum.}
\label{fig:LYsubfigs}
\end{figure}

Light yield measurements are performed with the dedicated experimental setup shown in Fig.~\ref{fig:LYsetup} housed in a custom-made dark box. Liquid scintillator samples of constant volume in a standard vial are placed in a fixed sample holder above a high quantum efficiency (peak quantum efficiency 35\%) 1.5" diameter photomultiplier (Hamamatsu R9420-100). A 10 $\mu$Ci $^{137}$Cs source is used to irradiate the scintillator from a fixed location, producing predominantly Compton electrons that deposit energy in the scintillator volume. Photomultiplier signals are collected and amplified before collecting a spectrum with a multi-channel analyser (FastComTec). Since the distribution of energy deposits between samples should be identical, for a fixed scintillation light yield, the measured spectra should also be identical. This can be quantified by comparing spectra measured under the same conditions using an analysis inspired by Ref.~\cite{Bignell_2015} and illustrated in Fig.~\ref{fig:LYdemo}. One spectrum is taken as a fixed reference, while the other is scaled by factors to emulate the spectrum that would be collected were the light yield higher or lower. The scaling factor which minimises the $\chi^2$ between the two spectra, calculated over a pre-defined spectral range, is taken as the `best fit' relative light yield. Repeated measurements of a single scintillator sample confirm that we can reproducibly measure light yields to 1\%.

\subsubsection{Compatibility Tests}
While a number of characterisation studies relating to the SABRE veto scintillator are ongoing, we report here on material compatibility tests that have been underway for over 100 days at the time of writing. The SABRE veto design includes numerous materials that will be in contact with the liquid scintillator over the multi-year operating life of the experiment. To ensure chemical compatibility, tests are underway using material samples submerged in a LAB-based test scintillator. The test scintillator was composed of purified LAB, 3 g/L PPO, and 15 mg/L bis-MSB (see Section~\ref{sec:LS}). 

A summary of the tested materials is given in Table~\ref{tab:Compatibility}. One important question considered in these tests was whether the steel vessel would be fully compatible with the LAB scintillator. The steel vessel represents the largest surface area in contact with the LAB out of any material, so to ensure its compatibility we considered the possibility of coating with a chemically inert fluoropolymer. Many of the remaining materials in Table~\ref{tab:Compatibility} relate to the potted bases of the veto PMTs that will be required to allow the high voltage and detector signal cables to be used when the PMTs are immersed in LAB.

\begin{table}[!t]
\centering
\caption{The materials used in the veto scintillator compatibility tests. Abbreviations: polytetrafluoroethylene (PTFE), fluoroelastopolymer (FEP), polyolefin (PO), photomultiplier tube (PMT).}
\label{tab:Compatibility}       
\begin{tabular}{lll}
\hline
Label & Material & Relevance  \\\hline
SS & Stainless steel & Veto vessel \\
SS-PTFE & PTFE-coated steel & Vessel coating \\
SS-FEP & FEP-coated steel & Vessel coating \\
Stud & Steel stud bolt & PMT mount point \\
Viton & Viton o-ring & Vessel flange seals \\
BlackCable & Coaxial cable & veto PMT \\
Epoxy & Epoxy sealant & veto PMT base \\
Potting & Potting compound & veto PMT base \\
FEPtube & FEP heat-shrink & veto PMT base \\
POtube & PO heat-shrink & veto PMT base \\
\hline
\end{tabular}
\end{table}

Two samples of each material were placed in contact with the scintillator inside glass bottles capped with a PTFE-lined lid and left in an air-conditioned room. Several control samples with no material in contact with the scintillator were also prepared. For the first 10 weeks of the compatibility tests, weekly measurements of the light yield relative to the control and the optical attenuation length relative to the control, were performed on one of these samples. The second `reserve' sample of each material was not measured in order to control for contamination during the measurement process. After 10 weeks, a comparison was made between the two samples, and the measurement cadence shifted to monthly.

\begin{figure}
    \centering
    \includegraphics[width=8cm]{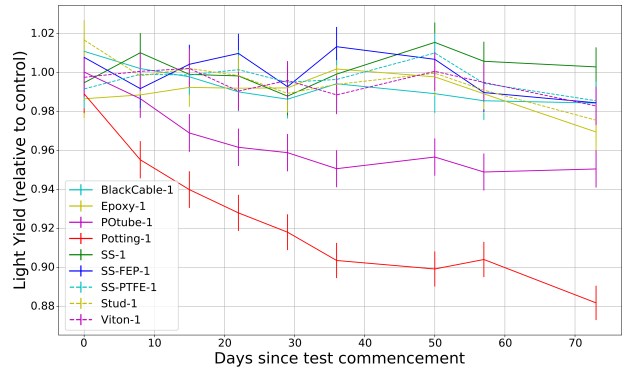}
    \caption{The light yield of all liquid scintillator samples placed in contact with materials to be submerged in the liquid scintillator veto, relative to a control with no material present.}
    \label{fig:LYcompatibility}
\end{figure}

\begin{figure}
    \centering
    \includegraphics[width=8cm]{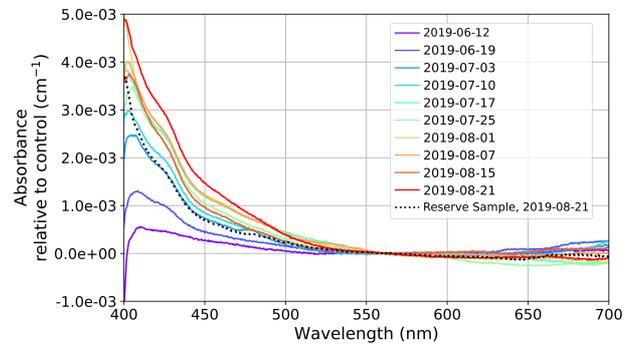}
    \caption{The optical absorbance of the liquid scintillator containing submerged polyolefin heat-shrink tube, relative to a control with no material. The legend labels denote the measurement date.}
    \label{fig:UVVIScompatibility}
\end{figure}

Two of the materials (potting compound and polyolefin heat shrink) were observed to affect the light yield (see Fig.~\ref{fig:LYcompatibility}), with the reserve samples confirming the results within uncertainty. Only one material (polyolefin heat shrink) was observed to significantly affect the optical absorbance (see Fig.~\ref{fig:UVVIScompatibility}). The reserve polyolefin tube sample also exhibited degradation at a somewhat lower level, although it is not clear whether the excess in the measured sample was due to contamination or normal variation between the polyolefin samples.

The polyolefin tube will not be used in the potted PMT base design for SABRE. The potting compound will be used, but is sealed inside an acrylic box with the tested epoxy compound and thus should not be directly exposed to the scintillator. A key outcome from these tests was the confirmation that the veto vessel need not be coated to ensure compatibility; if any degradation scales as the surface area to volume ratio and scales linearly with time in contact with the material, then the lack of observable change (at the 1\% level) after 100 days in our test geometry, translates to a stability at the level of 0.6\% over the 3-year life of the experiment.

\section{Future Work}
We have given an overview of detector characterisation studies that have been done to date for SABRE at the ANU. A more complete analysis of the quenching factors with Monte Carlo calculations of the expected distribution of recoil energies will be included in a future publication. 
The quenching factor measurements also provide a labelled dataset of nuclear recoils and electron recoils at low energy in NaI:Tl. Studies that use this dataset are underway to investigate particle identification approaches in NaI:Tl. Effective particle identification can help to further suppress backgrounds for SABRE.
The LAB compatibility tests are ongoing, and a new measurement campaign with further materials to be submerged in the veto liquid will commence shortly.

\begin{acknowledgement}
The authors thank A. Duffy and S. Krishnan for use of the NaI:Tl crystal from Swinburne University of Technology, and the ANU technical staff for their support preparing the beamline for the quenching factor measurements. I.M. and F.N acknowledge travel support from AINSE for their attendance at the ANU quenching factor measurements. This research was supported by the Australian Research Council, grant numbers DP170101675, LE160100080, LE170100162 and LE190100196. 
\end{acknowledgement}

\bibliography{library}

\begin{thebibliography}{27}

\bibitem{Klasen_2015}
M.~Klasen, M.~Pohl, G.~Sigl, Progress in Particle and Nuclear Physics
  \textbf{85}, 1–32 (2015)

\bibitem{Freese_2013}
K.~Freese, M.~Lisanti, C.~Savage, Reviews of Modern Physics \textbf{85},
  1561–1581 (2013)

\bibitem{Bernabei2018}
R.~Bernabei, P.~Belli, A.~Bussolotti, F.~Cappella, V.~Caracciolo, R.~Cerulli,
  C.~Dai, A.~d’ Angelo, A.~Di~Marco, et~al., Nuclear Physics and Atomic
  Energy \textbf{19}, 307–325 (2018)

\bibitem{Davis_2014}
J.H. Davis, Physical Review Letters \textbf{113} (2014)

\bibitem{mckinsey2018dama}
D.N. McKinsey, arXiv \textbf{[hep-ex]}, 1803.10110 (2018)

\bibitem{Adhikari_2019_modulation}
G.~Adhikari, P.~Adhikari, E.B. de~Souza, N.~Carlin, S.~Choi, M.~Djamal,
  A.~Ezeribe, C.~Ha, I.~Hahn, E.~Jeon et~al., Physical Review Letters
  \textbf{123} (2019)

\bibitem{Amare_2019_modulation}
S.~Amar\'e, S.~Cebri\'an, I.~Coarasa, C.~Cuesta, E.~Garc\'ia, M.~Mart\'inez,
  M.~Oliv\'an, Y.~Ortigoza, A.~Ortiz~de Sol\'orzano, K.~Puimed\'on et~al.,
  Physical Review Letters \textbf{123} (2019)

\bibitem{Ohare2015}
C.A.J. O'Hare, A.M. Green, J.~Billard, E.~Figueroa-Feliciano, L.E. Strigari,
  Phys. Rev. D \textbf{92}, 063518 (2015)

\bibitem{urquijo2016southern}
P.~Urquijo, arXiv \textbf{[physics.ins-det]}, 1605.03299 (2016)

\bibitem{BernabeiConf2018}
{Bernabei, R.}, {Belli, P.}, {Cappella, F.}, {Caracciolo, V.}, {Cerulli, R.},
  {Dai, C.J.}, {d'Angelo, A.}, {Di Marco, A.}, {He, H.L.}, {Incicchitti, A.}
  et~al., EPJ Web Conf. \textbf{182}, 02027 (2018)

\bibitem{CoEDMPP}
\emph{{ARC} {C}entre of {E}xcellence for {D}ark {M}atter {P}article {P}hysics},
  \url{https://www.darkmatter.org.au/}, accessed: 2019-11-01

\bibitem{PDG2019}
M.~Tanabashi, K.~Hagiwara, K.~Hikasa, K.~Nakamura, Y.~Sumino, F.~Takahashi,
  J.~Tanaka, K.~Agashe, G.~Aielli, C.~Amsler et~al. (Particle Data Group),
  Phys. Rev. D \textbf{98}, 030001 (2018)

\bibitem{Antonello_2019_MC}
M.~Antonello, E.~Barberio, T.~Baroncelli, J.~Benziger, L.~Bignell,
  I.~Bolognino, F.~Calaprice, S.~Copello, D.~D’Angelo, G.~D’Imperio et~al.,
  Astroparticle Physics \textbf{106}, 1–9 (2019)

\bibitem{Antonello_2019_PoP}
M.~Antonello, E.~Barberio, T.~Baroncelli, J.~Benziger, L.J. Bignell,
  I.~Bolognino, F.~Calaprice, S.~Copello, D.~D’Angelo, G.~D’Imperio et~al.,
  The European Physical Journal C \textbf{79} (2019)

\bibitem{Collar2010}
J.I. Collar, arXiv \textbf{[astro-ph.IM]}, 1010.5187 (2010)

\bibitem{BIRKS1964}
J.B. Birks, in \emph{The Theory and Practice of Scintillation Counting}
  (Pergamon, 1964), International Series of Monographs in Electronics and
  Instrumentation, pp. 68 -- 95, ISBN 978-0-08-010472-0

\bibitem{Collar2013}
J.I. Collar, Phys. Rev. C \textbf{88}, 035806 (2013)

\bibitem{Xu2015}
J.~Xu, E.~Shields, F.~Calaprice, S.~Westerdale, F.~Froborg, B.~Suerfu,
  T.~Alexander, A.~Aprahamian, H.O. Back, C.~Casarella et~al., Phys. Rev. C
  \textbf{92}, 015807 (2015)

\bibitem{stiegler2017QF}
T.~Stiegler, C.~Sofka, R.C. Webb, J.T. White, arXiv \textbf{[physics.ins-det]},
  1706.07494 (2017)

\bibitem{joo2018quenching}
H.W. Joo, H.S. Park, J.H. Kim, S.K. Kim, Y.D. Kim, H.S. Lee, S.H. Kim,
  \emph{Quenching factor measurement for nai(tl) scintillation crystal} (2018),
  \texttt{1809.10310}

\bibitem{ValentineAndRooney1998}
J.D. {Valentine}, B.D. {Rooney}, P.~{Dorenbos}, IEEE Transactions on Nuclear
  Science \textbf{45}, 1750 (1998)

\bibitem{Khodyuk_2010}
I.V. Khodyuk, P.A. Rodnyi, P.~Dorenbos, Journal of Applied Physics
  \textbf{107}, 113513 (2010)

\bibitem{XIA}
\emph{Xia llc}, \url{https://xia.com/}, accessed: 2019-12-01

\bibitem{smith2003DSP}
S.~Smith, \emph{Digital Signal Processing: A Practical Guide for Engineers and
  Scientists} (Elsevier Science, 2003), ISBN 9780750674447

\bibitem{Reifarth_2009}
R.~Reifarth, M.~Heil, F.~Käppeler, R.~Plag, Nuclear Instruments and Methods in
  Physics Research Section A: Accelerators, Spectrometers, Detectors and
  Associated Equipment \textbf{608}, 139–143 (2009)

\bibitem{Benziger_2008}
J.~Benziger, L.~Cadonati, F.~Calaprice, M.~Chen, A.~Corsi, F.~Dalnoki-Veress,
  R.~Fernholz, R.~Ford, C.~Galbiati, A.~Goretti et~al., Nuclear Instruments and
  Methods in Physics Research Section A: Accelerators, Spectrometers, Detectors
  and Associated Equipment \textbf{587}, 277–291 (2008)

\bibitem{Bignell_2015}
L.~Bignell, M.~Diwan, S.~Hans, D.~Jaffe, R.~Rosero, S.~Vigdor, B.~Viren,
  E.~Worcester, M.~Yeh, C.~Zhang, Journal of Instrumentation \textbf{10},
  P10027–P10027 (2015)

\end{thebibliography}
%
%

\end{document}